\documentclass[
reprint,amsmath,amssymb,aps,pra,floatfix,]{revtex4-1}
\usepackage{graphicx}	
\usepackage{dcolumn}	
\usepackage{bm}		
\usepackage{color}

\begin{document}

\title{Interference between two resonant transitions with 
distinct initial and final states connected by radiative decay 
}

\author{A. Marsman, M. Horbatsch and E.A. Hessels}
\email{hessels@yorku.ca}

\affiliation{%
Department of Physics and Astronomy, York University, Toronto, Ontario M3J 1P3, Canada
}%

\date{\today}

\begin{abstract}
The resonant line shape from driving 
a transition between two states, 
$|\rm{a}\rangle$ 
and 
$|\rm{b}\rangle$, 
can be distorted due to a quantum-mechanical interference
effect involving a resonance between two different states, 
$|\rm{c}\rangle$ 
and 
$|\rm{d}\rangle$,
if 
$|\rm{c}\rangle$ 
has a decay path to 
$|\rm{a}\rangle$ 
and 
$|\rm{d}\rangle$ 
has a decay path to 
$|\rm{b}\rangle$.
This interference can cause a shift of the
measured resonance, 
despite the fact that the two resonances do not have 
a common initial or final state.
As an example,
we demonstrate that
such a shift affects measurements of the 
atomic hydrogen 
2S$_{1/2}$-to-2P$_{1/2}$ 
Lamb-shift
transition 
due to  
3S-to-3P 
transitions if the 
3S$_{1/2}$ 
state has some 
initial population. 
\begin{description}
\item[PACS numbers]
\verb+\pacs{32.70.Jz,06.20.Jr}+
\end{description}
\end{abstract}

\maketitle

\section{introduction}

Recently, 
the effect of 
quantum-mechanical interference on precision measurements
has been investigated by the 
present authors 
\cite{
PRA.82.052519,
PRA.84.032508,
PRA.86.012510,
PRA.86.040501,
PRA.89.043403,
PRA.91.062506}
and by others 
\cite{
PRL.107.023001,
PRA.87.032504,
PRL.109.259901,
PhysRevA.92.022514,
PhysRevA.90.012512,
PhysRevA.92.062506,
PhysRevA.94.042111,
PhysRevA.95.052503,
PhysRevA.91.030502,
AccurateLineshape}.
These investigations indicate that interference 
with a neighboring resonance,
even if it is very distant, 
can lead to significant shifts for precision measurements.
All of the investigations involve the interference between 
a pair of resonances that have a common initial state.
Here we investigate the interference between two 
resonances that do not have a common initial state,
nor a common final state.
We find that an interference between two such resonances 
exists if the four states are connected by radiative decay, 
and this interference causes a shift
of the measured resonance center.

We explore the implications of this type of effect on 
the precision measurement of the atomic hydrogen 
2S$_{1/2}$-to-2P$_{1/2}$ 
Lamb shift,
which was measured 
by Lundeen and Pipkin 
\cite{PRL46.232}
some time ago,
and is currently being
remeasured by our group.
We find that 
a shift in this measured resonance can be present 
if there is an initial population in the 
3S$_{1/2}$ 
state,
due to interference between the 
2S$_{1/2}$-to-2P$_{1/2}$ 
resonance and the
3S$_{1/2}$-to-3P 
resonances.
The hydrogen Lamb-shift measurement has become 
important since it can,
when compared to very precise theory 
\footnote{for an overview of measurements of the 
hydrogen Lamb shifts and 
theoretical predictions, 
see
Refs. 
\cite{horbatsch2016tabulation}
and
\cite{mohr2016codata}},
determine the charge radius of the proton. 
More precise determinations of this radius have now been
performed using muonic hydrogen 
\cite{
pohl2010size,
antognini2013proton},
but there is a large discrepancy between measurements
made using ordinary hydrogen and muonic hydrogen.
This discrepancy has become known as the proton size 
puzzle 
\cite{
bernauer2014proton,
pohl2013muonic,
carlson2015proton}.

\section{Four-level system \label{sect:4Level}}

\begin{figure}
\includegraphics[width=1.5in]{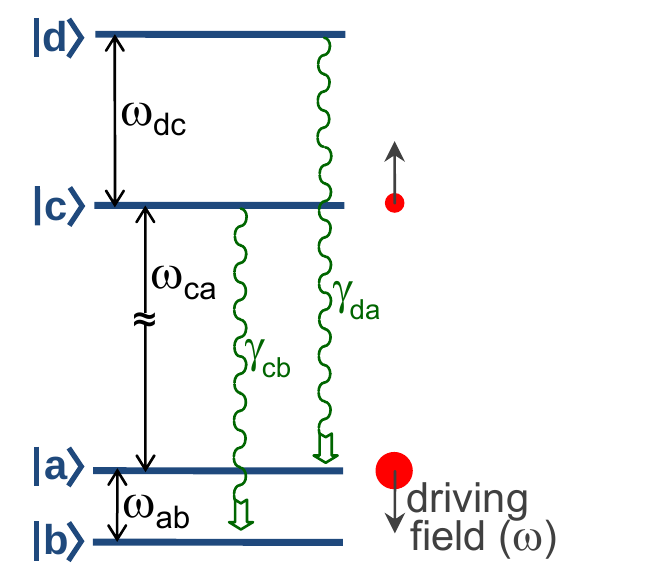}
\caption{\label{fig:4levels} (Color online) 
Energy-level diagram for a four-level system. 
States 
$|\rm{c}\rangle$
and
$|\rm{d}\rangle$
radiatively decay to 
$|\rm{b}\rangle$
and
$|\rm{a}\rangle$,
respectively.
When driving the 
$|\rm{a}\rangle$-to-$|\rm{b}\rangle$
transition,
any population in 
$|\rm{c}\rangle$
can be driven
(by far-off-resonance radiation)
to 
$|\rm{d}\rangle$.
An interference between the 
$|\rm{a}\rangle$-to-$|\rm{b}\rangle$
and
$|\rm{c}\rangle$-to-$|\rm{d}\rangle$
transitions,
as described in the text,
leads to a distortion of the line shape,
and a shift of the  
$|\rm{a}\rangle$-to-$|\rm{b}\rangle$
resonance line center.
}
\end{figure}

We first consider a four-level system consisting of
states 
$|\rm{a}\rangle$,
$|\rm{b}\rangle$,
$|\rm{c}\rangle$,
and $|\rm{d}\rangle$,
as shown in 
Fig.~\ref{fig:4levels}.
An electric field 
$\vec{E}(t)=E_0 \hat{z} \cos(\omega t +\phi)$
drives the 
$|\rm{a}\rangle$-to-$|\rm{b}\rangle$
electric-dipole 
transition, 
depleting the initially populated 
$|\rm{a}\rangle$ state. 
The two other states,
$|\rm{c}\rangle$,
and $|\rm{d}\rangle$,
are also assumed to be connected by an
electric-dipole matrix element, 
but the transition between these two states is far out of
resonance with 
$\vec{E}(t)$.
Since 
$|\rm{d}\rangle$  
decays radiatively to 
$|\rm{a}\rangle$, 
any initial population in 
$|\rm{c}\rangle$
that is weakly driven by the off-resonant field to 
$|\rm{d}\rangle$
slightly enhances the population of the
$|\rm{a}\rangle$ 
state. 
A more efficient method of 
transferring population from 
$|\rm{c}\rangle$
to
$|\rm{a}\rangle$
involves an interference between the 
on-resonant 
$|\rm{a}\rangle$-to-$|\rm{b}\rangle$
transition amplitude
and the off-resonant
$|\rm{c}\rangle$-to-$|\rm{d}\rangle$
transition, 
which is possible because of the radiative
decay from a coherence between 
$|\rm{c}\rangle$
and 
$|\rm{d}\rangle$
to a coherence between
$|\rm{a}\rangle$
and
$|\rm{b}\rangle$.
The relevant equations for the density matrix
for the four-level system are
\begin{subequations}
\begin{eqnarray}
\dot\rho_{\rm aa}&=& 
i V_{\rm ab}^* \rho_{\rm ab} - i V_{\rm ab} \rho_{\rm ba}  
+ \gamma_{\rm da} \rho_{\rm dd},
\\
\dot\rho_{\rm bb}&=& 
i V_{\rm ab} \rho_{\rm ba} - i V_{\rm ab}^* \rho_{\rm ab}  
+ \gamma_{\rm cb} \rho_{\rm cc},
\\
\dot\rho_{\rm ba}&=& 
i \omega_{\rm ab} \rho_{\rm ba}
+ i V_{\rm ab}^* (\rho_{\rm bb}-\rho_{\rm aa}) 
\pm \sqrt{\gamma_{\rm da} \gamma_{\rm cb}} \rho_{\rm cd},
\\
\dot\rho_{\rm cc}&=& 
i V_{\rm cd}^* \rho_{\rm cd} - i V_{\rm cd} \rho_{\rm dc}  
- \gamma_{\rm cb} \rho_{\rm cc},
\\
\dot\rho_{\rm dd}&=& 
i V_{\rm cd} \rho_{\rm dc} - i V_{\rm cd}^* \rho_{\rm cd}  
- \gamma_{\rm da} \rho_{\rm dd},
\\
\dot\rho_{\rm cd}&=& 
i \omega_{\rm dc} \rho_{\rm cd} 
+ i V_{\rm cd} (\rho_{\rm cc}-\rho_{\rm dd}) 
-\frac{\gamma_{\rm da} + \gamma_{\rm cb}}{2} \rho_{\rm cd},\ \ 
\end{eqnarray}
\label{eq:4LevelEqs}
\end{subequations}
where 
$V_{ij}(t)=
\langle i|\ e \vec{E}(t) \cdot \vec{r}\ |j \rangle / \hbar$,
and 
Fig.~\ref{fig:4levels}
shows the energy separations,
$\hbar \omega_{ij}$,
and 
decay rates,
$\gamma_{ij}$.
Of particular interest in these equations is the 
square-root
term, 
which allows for the coherence between
$|\rm{c}\rangle$
and
$|\rm{d}\rangle$
to be transferred via radiative decay to a
coherence between
$|\rm{a}\rangle$
and
$|\rm{b}\rangle$.
This term 
\cite{PRA.27.2456} 
is often incorrectly omitted in the treatment of 
density matrices. 
The sign of the term depends on the 
relative sign of the 
a-to-d 
and 
b-to-c
matrix elements
that are responsible for the radiative decay.

\begin{figure}
\includegraphics[width=3.2in]{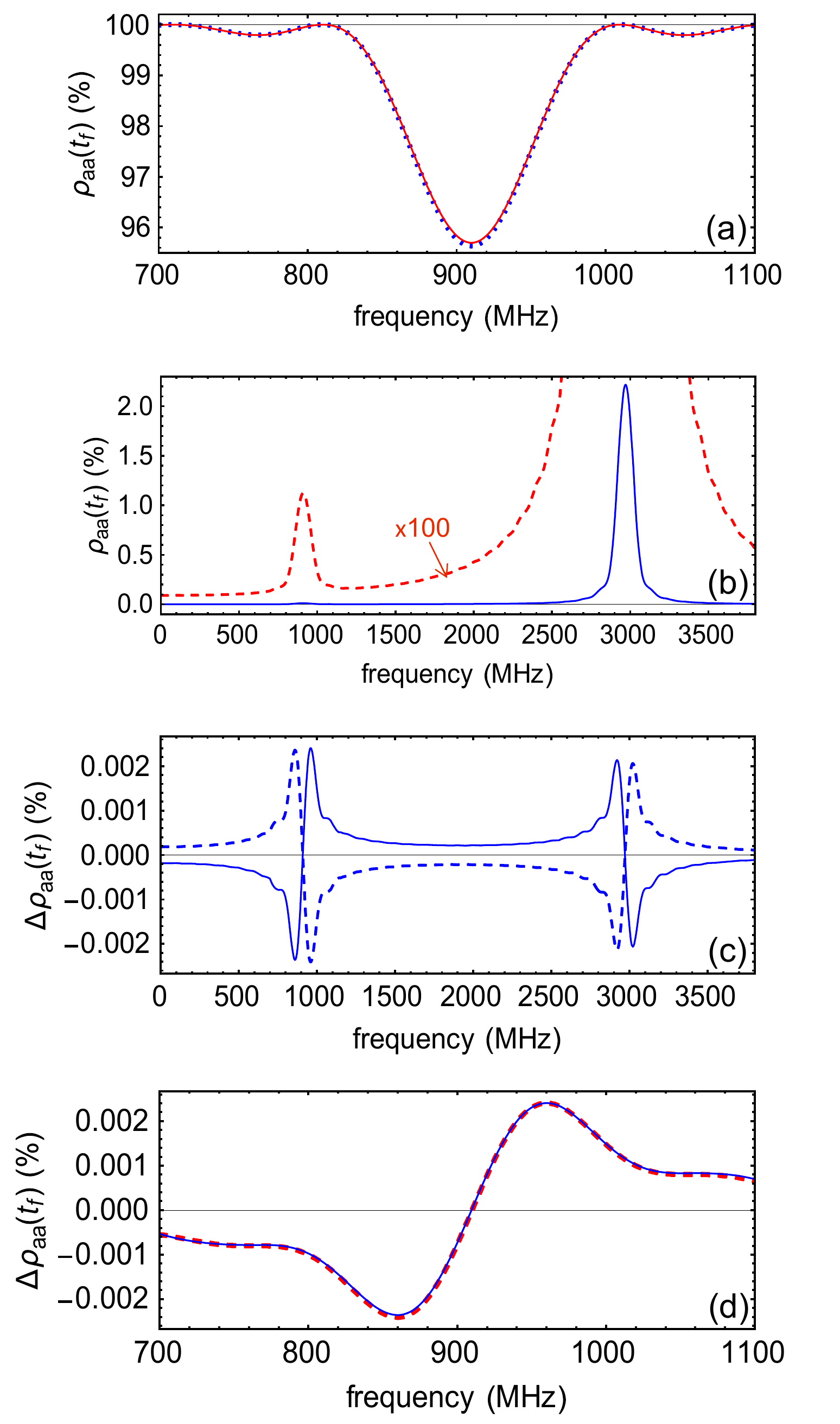}
\caption{\label{fig:RabiLineShapes} (Color online) 
Line shapes for a four-level system with 
initial population in 
$|\rm{a}\rangle$ 
(a), 
and for initial population in 
$|\rm{c}\rangle$ 
(b, c, and d). 
The solid, red line in 
(a) 
is the full calculation, 
with the dotted, blue line showing 
the perturbative approximation of
Eq.~(\ref{eq:PertLineShape}).
The larger resonance in  
(b)
is from driving the 
$|\rm{c}\rangle$-to-$|\rm{d}\rangle$
transition, followed by decay down to
$|\rm{a}\rangle$.
The smaller resonance 
(expanded by a factor of 100 in the 
dashed curve)
results from radiative decay to 
$|\rm{b}\rangle$, 
followed by driving of the 
$|\rm{b}\rangle$-to-$|\rm{a}\rangle$
transition.
The line shape in 
(b) 
artificially suppresses the effect of the 
coherent decay term 
(the 
square-root 
term in 
Eq.~(\ref{eq:4LevelEqs}c)).
The effect of this term is shown in 
(c),
with the solid and dashed lines
showing the effects for the positive
and negative signs of the term.
This distortion,
for the positive case,
is shown on a larger 
scale in 
(d),
along with the perturbative 
approximation of 
Eq.~(\ref{eq:PertApproxEq}).
When measuring the line center of the
resonance in 
(a), 
a population in state 
$|\rm{c}\rangle$ 
can cause a shift in this resonance
due to the dispersion line shape of 
(c). 
For these plots, 
the parameters used 
(See Eq.~(\ref{eq:4LevelEqs})
and Fig.~\ref{fig:4levels})
are:
$\omega_{\rm ba}$=2$\pi$(909.874~MHz),
$\omega_{\rm dc}$=2$\pi$(2$\,$970.292~MHz),
$\gamma_{\rm da}$=14.949~MHz,
$\gamma_{\rm cb}$=0.701~MHz,
$V_{\rm ab}$=13.925$\frac{\rm MHz}{\rm V/cm} E(t)$,
$V_{\rm cd}$=$-48.238$$\frac{\rm MHz}{\rm V/cm} E(t)$,
$t_f$=10~ns,
and 
$E_0$=3~V/cm.
}
\end{figure}

Figure~\ref{fig:RabiLineShapes}
shows the results of integrating 
Eq.~(\ref{eq:4LevelEqs}) 
from
$t$=0
to 
$t$=$t_f$
for one choice of the 
parameters of
Eq.~(\ref{eq:4LevelEqs}),
with an initial
population in 
$|\rm{a}\rangle$
($\rho_{\rm aa}(0)$=1,
Fig. ~\ref{fig:RabiLineShapes}(a)) 
and with an initial
population in 
$|\rm{c}\rangle$
($\rho_{\rm cc}(0)$=1,
Fig.~\ref{fig:RabiLineShapes}(b) and (c)).
The line shapes are averaged over the
phase 
$\phi$
of the driving field.
An initial population in 
$|\rm{a}\rangle$
leads to a simple resonance at 
$f$=$\frac{\omega_{\rm ab}}{2 \pi}$,
as shown in 
Fig.~\ref{fig:RabiLineShapes}(a).
For the small value of 
$E_0$ 
used,
it is well approximated by the 
perturbative expression 
\begin{equation}
\rho_{\rm aa}(t_f)=1-
\frac{V_{\rm ab}^2 \sin^2(\Delta\omega\  t_f/2)}
{\Delta\omega^2},
\label{eq:PertLineShape}
\end{equation}
where
$\Delta\omega=\omega - \omega_{\rm ab}$.

An examination of 
Eq.~(\ref{eq:4LevelEqs})
reveals that the three lowest-order routes 
for an initial population in 
$|\rm{c}\rangle$ 
to get to state
$|\rm{a}\rangle$
are
\begin{subequations}
\begin{eqnarray}
&&\rho_{\rm cc} \to \rho_{\rm cd},\, \rho_{\rm dc} \to \rho_{\rm dd} 
\to \rho_{\rm aa},
\\
&&\rho_{\rm cc} \to \rho_{\rm bb} \to \rho_{\rm ba}, \,\rho_{\rm ab} 
\to \rho_{\rm aa},
\\
\rm{and}\  &&\rho_{\rm cc} \to \rho_{\rm cd}, \,\rho_{\rm dc} 
\to \rho_{\rm ba},\, \rho_{\rm ab} \to \rho_{\rm aa}.
\end{eqnarray}
\label{eq:3pathsEqs}
\end{subequations}
Eq.~(\ref{eq:3pathsEqs}a)
leads to a resonance centered at 
$f$=$\frac{\omega_{\rm dc}}{2 \pi}$, 
and 
such a resonance is clearly seen 
in Fig.~\ref{fig:RabiLineShapes}(b).
When the scale is expanded by a factor of 100 
(dashed line in 
Fig.~\ref{fig:RabiLineShapes}(b)),
the smaller resonance at 
$f$=$\frac{\omega_{\rm ab}}{2 \pi}$
caused by
Eq.~(\ref{eq:3pathsEqs}b)
is visible.

The result shown in 
Fig.~\ref{fig:RabiLineShapes}(b)
is calculated by artificially suppressing the 
square-root 
term in 
Eq.~(\ref{eq:4LevelEqs}c).
The additional effect of this term is shown in 
Fig.~\ref{fig:RabiLineShapes}(c),
with the solid curve showing the result
for the positive sign of this term and the 
dashed curve for the negative sign.
These curves show the effect due to
Eq.~(\ref{eq:3pathsEqs}c). 
For the small values of the 
$V$ 
and 
$\gamma$
parameters used for 
Fig.~\ref{fig:RabiLineShapes},
the effect of the 
square-root 
term can be approximated 
by the perturbative expression
\begin{equation}
\Delta\rho_{\rm aa}(t_f)=
\frac{\pm 
\sqrt{\gamma_{\rm cb} \gamma_{\rm da} }\  t_f V_{\rm ab} V_{\rm cd} }
{2 \Delta\omega (\omega_{\rm dc} - \omega_{\rm ab} ) }
\left( 1 -
\frac{\sin(\Delta\omega\  t_f)}
{\Delta\omega\  t_f}
\right).
\label{eq:PertApproxEq}
\end{equation}
A comparison 
(for the positive choice of sign) 
of this perturbative expression to the 
full result of 
Fig.~\ref{fig:RabiLineShapes}(c)
is shown in 
Fig.~\ref{fig:RabiLineShapes}(d).

Figure~\ref{fig:RabiLineShapes}
shows that the measurement of the line center of 
the 
$|\rm{a}\rangle$-to-$|\rm{b}\rangle$
resonance can be affected by a population in 
$|\rm{c}\rangle$.
A small distortion and shift of the 
$|\rm{a}\rangle$-to-$|\rm{b}\rangle$
resonance results from the long tail of 
the resonance at 
$\frac{\omega_{\rm dc}}{2 \pi}$
(as seen by the dashed line in 
Fig.~\ref{fig:RabiLineShapes}(b)),
which leads to a sloped background under the 
$|\rm{a}\rangle$-to-$|\rm{b}\rangle$
resonance of 
Fig.~\ref{fig:RabiLineShapes}(a).
If 
$|\rm{a}\rangle$
and
$|\rm{c}\rangle$
start with equal populations, 
the effect of this slope is to shift the
$|\rm{a}\rangle$-to-$|\rm{b}\rangle$
resonance by only
0.4 kHz
for the parameters used in the figure.

A larger shift results from the effect of 
Eq.~(\ref{eq:3pathsEqs}c)
(Fig.~\ref{fig:RabiLineShapes}(c) and (d)).
The dispersion line shape due to this
interference effect, 
causes a shift of 33 kHz for the parameters
used in 
Fig.~\ref{fig:RabiLineShapes}.
This shift is independent of 
$E_0$
for small values of 
$E_0$, 
since 
Fig.~\ref{fig:RabiLineShapes}(a)
and 
Fig.~\ref{fig:RabiLineShapes}(c)
both scale as $E_0^2$ below saturation,
as can be seen from 
Eqs.~(\ref{eq:PertLineShape})
and (\ref{eq:PertApproxEq}).
Examination of these equations at 
$\Delta\omega$=$\pm \pi/t_f$ 
(points near half maximum)
shows that the shift due to 
the effect of the interference can be 
approximated by
\begin{equation}
\Delta f =
\mp
\frac{\pi}{8}
\frac{1}{t_f}
\frac{V_{\rm cd}}{V_{\rm ab}}
\frac
{\sqrt{\gamma_{\rm cb} \gamma_{\rm da} }}
{(\omega_{\rm dc} - \omega_{\rm ab}) }
\frac{\rho_{\rm cc}(t\!=\!0)}{\rho_{\rm aa}(t\!=\!0)}.
\label{eq:shiftApprox}
\end{equation}
The result is a shift that persists,
even when extrapolated to zero intensity
for the driving field.
For higher
$E_0$, 
the shift becomes 
somewhat larger since the resonance
being measured saturates, 
while the distortion of 
Fig.~\ref{fig:RabiLineShapes}(c) 
continues to increase.

\section{Separated Oscillatory Fields \label{SOF}}

Shifts are also found when the method of 
separated oscillatory fields
(SOF) 
\cite{PhysRev.76.996} 
is used for the measurement. 
For SOF measurements, 
a field of amplitude 
$E_0$
and phase
$\phi$
is present from 
$t$=0
to
$t$=$t_D$,
and a second field of the 
same amplitude that
is either in phase or out of phase
by 
180 degrees
($\phi$
or
$\phi+\pi$)
is present from 
$t$=$t_S$
to
$t$=$t_S+t_D$.
The SOF line shape is the difference between the 
in-phase and out-of-phase signals.

\begin{figure} [t]
\includegraphics[width=3.2in]{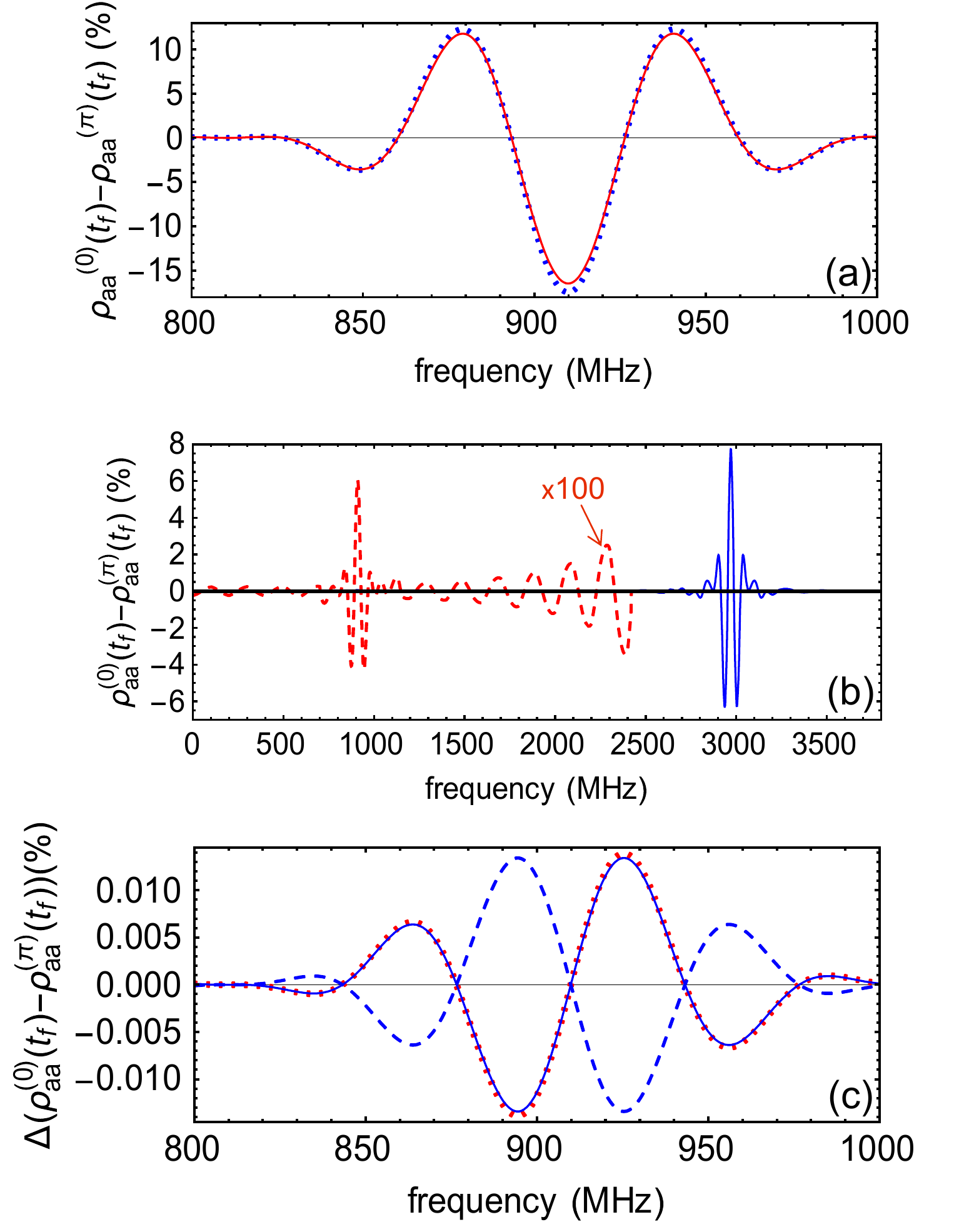}
\caption{\label{fig:SOFLineShapes} (Color online) 
Line shapes for a four-level system 
using separated oscillatory fields.
Initial population in 
$|\rm{a}\rangle$ 
leads to a typical SOF line shape,
(a), 
that is well approximated by 
a perturbative expression 
(dotted line).
Initial population in 
$|\rm{c}\rangle$ 
(b and c)
leads to features 
due to the 
$|\rm{c}\rangle$-to-$|\rm{d}\rangle$
transition
(the large feature in (b)),
due to the 
$|\rm{b}\rangle$-to-$|\rm{a}\rangle$
transition
(the smaller feature in (b), 
expanded by a factor of 100 in the 
dashed curve),
and an interference between these
two resonances, 
as shown in 
(c) with both the positive
(solid blue line) 
and negative
(dashed blue line)
sign choices in 
Eq.~(\ref{eq:4LevelEqs}c).
The perturbative approximation 
(Eq.~(\ref{eq:PertApproxSOF}))
for the positive sign is shown
by the dotted red line.
When measuring the line center of the
resonance in 
(a), 
a population in state 
$|\rm{c}\rangle$ 
can cause a shift in this resonance
due to the dispersion line shape of 
(c). 
For these plots, 
the parameters used are as in 
Fig.~\ref{fig:RabiLineShapes},
with
$t_D$=10~ns,
and 
$t_S$=15~ns.
}
\end{figure} 

Integrating 
Eq.~(\ref{eq:4LevelEqs})
from 
$t$=0
to
$t$=$t_f$=$t_D+t_S$
for the 
SOF 
fields 
for one choice of the parameters 
leads to the line shapes of
Fig.~\ref{fig:SOFLineShapes}.
For an initial population in
$|\rm{a}\rangle$,
Fig.~\ref{fig:SOFLineShapes}(a)
shows that the final population 
exhibits a typical 
SOF 
line shape.
This line shape is well approximated by its
perturbative expression
\begin{equation}
\frac{-4 V_{\rm ab}^2 \sin^2(\Delta\omega\  t_D/2)}
{\Delta\omega^2}
\cos(\Delta\omega\  t_S),
\label{eq:PertSOFShape}
\end{equation}
as shown by the dotted line.
 
The line shape is distorted if there is 
any starting population in 
$|\rm{c}\rangle$.
The large feature centered at
$\omega=\omega_{\rm dc}$
in 
Fig.~\ref{fig:SOFLineShapes}(b)
is due to 
Eq.~(\ref{eq:3pathsEqs}a),
and the much smaller feature centered at
$\omega=\omega_{\rm ab}$
results from 
Eq.~(\ref{eq:3pathsEqs}b).
The more important feature is due to
Eq.~(\ref{eq:3pathsEqs}c),
and this interference feature is shown
separately in 
Fig.~\ref{fig:SOFLineShapes}(c),
for both the positive 
(solid)
and negative
(dashed)
choices of sign in 
Eq.~(\ref{eq:4LevelEqs}c).
The perturbative expression
\begin{equation}
\mp 4 \sqrt{\gamma_{\rm cb} \gamma_{\rm da} } 
V_{\rm ab} V_{\rm cd}
\frac{ 
\sin^2(\Delta\omega\ t_D/2)  }
{\Delta\omega^2 (\omega_{\rm dc} - \omega_{\rm ab} ) }
\sin(\Delta\omega\ t_S),
\label{eq:PertApproxSOF}
\end{equation}
is a good approximation for the distortion for small
$E_0$,
as shown by the dotted red line in 
Fig.~\ref{fig:SOFLineShapes}(c).

The shifts caused by population in 
$|\rm{c}\rangle$ can be estimated
using the central zero-crossing positions in 
Fig.~\ref{fig:SOFLineShapes}(a).
Using this method, 
the effects of 
Eq.~(\ref{eq:3pathsEqs}a)
and 
(\ref{eq:3pathsEqs}b)
(Fig.~\ref{fig:SOFLineShapes}(b))
lead to a shift of +1.3 kHz
for the parameters used in the figure,
and the effect of the interfering term,
Eq.~(\ref{eq:3pathsEqs}c)
(Fig.~\ref{fig:SOFLineShapes}(c)),
leads to a shift of $\mp$9.3 kHz.
From an examination of 
Eqs.~(\ref{eq:PertSOFShape})
and
(\ref{eq:PertApproxSOF}),
it can be seen that this shift is given by
\begin{equation}
\Delta f =
\mp
\frac{1}{\pi}
\frac{1}{2 t_S}
\frac{V_{\rm cd}}{V_{\rm ab}}
\frac
{\sqrt{\gamma_{\rm cb} \gamma_{\rm da} }}
{(\omega_{\rm dc} - \omega_{\rm ab}) }
\frac{\rho_{\rm cc}(t=0)}{\rho_{\rm aa}(t=0)}.
\label{eq:SOFshiftApprox}
\end{equation}
As in Section~\ref{sect:4Level}, 
for small 
$E_0$,
the shifts are independent of 
field strength,
and therefore the shifts persist even
in the limit of zero field.
Both 
Eq.~(\ref{eq:shiftApprox})
and
Eq.~(\ref{eq:SOFshiftApprox})
show that the scale of the shift is
\begin{equation}
\frac{V_{\rm cd}}{V_{\rm ab}}
\,
\frac
{\sqrt{\gamma_{\rm cb} \gamma_{\rm da} }}
{(\omega_{\rm dc} \! - \! \omega_{\rm ab}) }
\,
\frac{\rho_{\rm cc}(t\!=\!0)}{\rho_{\rm aa}(t\!=\!0)}
\label{eq:shiftsScale}
\end{equation}
times the width of the resonance.

\section{the H($n$=2) Lamb shift}

The interference effects described in the previous
sections will affect a precision measurement of the
$n$=2 Lamb shift, 
if any population is present in higher-$n$ states.
For hydrogen atoms created by charge exchange,
as in 
Ref.~\cite{PRL46.232}
and our ongoing measurement,
the 
3S 
state has the largest charge-exchange
cross section for 
states with 
$n$$>$2.
For a 50-keV proton beam, 
the ratio of the 
charge-exchange 
cross sections
\cite{PhysRevA.1.1424} 
for the  
3S$_{1/2}$ 
and 
2S$_{1/2}$ 
states is approximately
0.34.

To illustrate the effect of the previous
sections with a more concrete example,
we calculate the shift that would be caused
in the SOF measurement of the
2S$_{1/2}$-to-2P$_{1/2}$ 
Lamb-shift interval of 
Ref.~\cite{PRL46.232},
if the experiment starts with 
a relative 
3S-to-2S 
population 
ratio of 
0.34.
We emphasize that the population
in the 
3S 
state could be different for 
the actual measurement of 
Ref.~\cite{PRL46.232}
(for example, due to multiple
collisions during charge exchange),
and also that other 
high-$n$
states could shift the measured
resonance. 
The calculation here,
however, 
sets a scale for possible shifts.

The calculation performed includes 
all states for 
$n$=1, 2, and 3.
These include all hyperfine states
(all values of quantum number 
$f$)
and all sublevels
(all values of 
$m_f$), 
for a total of 56 states.
Thus, 
Eq.~(\ref{eq:4LevelEqs})
must be expanded to 
$56^2$=3136 
equations. 
These equations have been
numerically integrated through
a trajectory that takes 
400~ns 
to traverse.
The fields along this trajectory
are those identified as configuration~1
in 
Ref.~\cite{LundeenMetrologia}.
As suggested in that reference,
a trajectory that is 
1.265~mm 
from the axis is used
(as this distance is the 
root-mean-squared
distance from the axis for the extended
beam of atoms in that work).
The 
off-axis 
trajectory leads
to two components of 
the electric field, 
with the largest component driving
$\Delta m_f$=0
transitions 
and the smaller component
driving
$\Delta m_f$=$\pm 1$
transitions.
The fields do not turn on and off
suddenly, 
as was assumed in 
Section~\ref{SOF},
but rather follow the more realistic
profiles described in 
Ref.~\cite{LundeenMetrologia}.

The
400-ns
trajectory includes
a 
35-ns-long 
preparation field at 
1110~MHz
(which depletes the 
2S $f$=1
population),
two 
10-ns-long
fields to perform the 
SOF
measurement
of the 
2S$_{1/2}$($f$=0)-to-2P$_{1/2}$($f$=1)
transition
(with the centers of these fields 
separated by 
15~ns),
and a 
10-ns-long 
field at 
910~MHz
used to quench the remaining
2S 
population
by mixing the 
2S 
state with the 
quickly-decaying
2P 
state.
As in 
Ref.~\cite{LundeenMetrologia},
the amplitudes used for
these fields are
26, 
11.4, 
and 
11~V/cm, 
respectively.
As in the experiment, 
the 
$n$=2-to-$n$=1 
radiative decay 
(Lyman-$\alpha$ 
fluorescence)
is monitored during the quench
field.
The calculation is repeated for 
different phases of each of these
fields, 
and averaged over these phases,
as well as taking the difference 
between the in-phase and 
180-degree-out-of-phase
signals for the SOF fields. 
The signals
(based on 
Lyman-$\alpha$
fluorescence induced by the quench fields)
are calculated for a range of frequencies
for the SOF fields.

\begin{figure}
\includegraphics[width=3.2in]{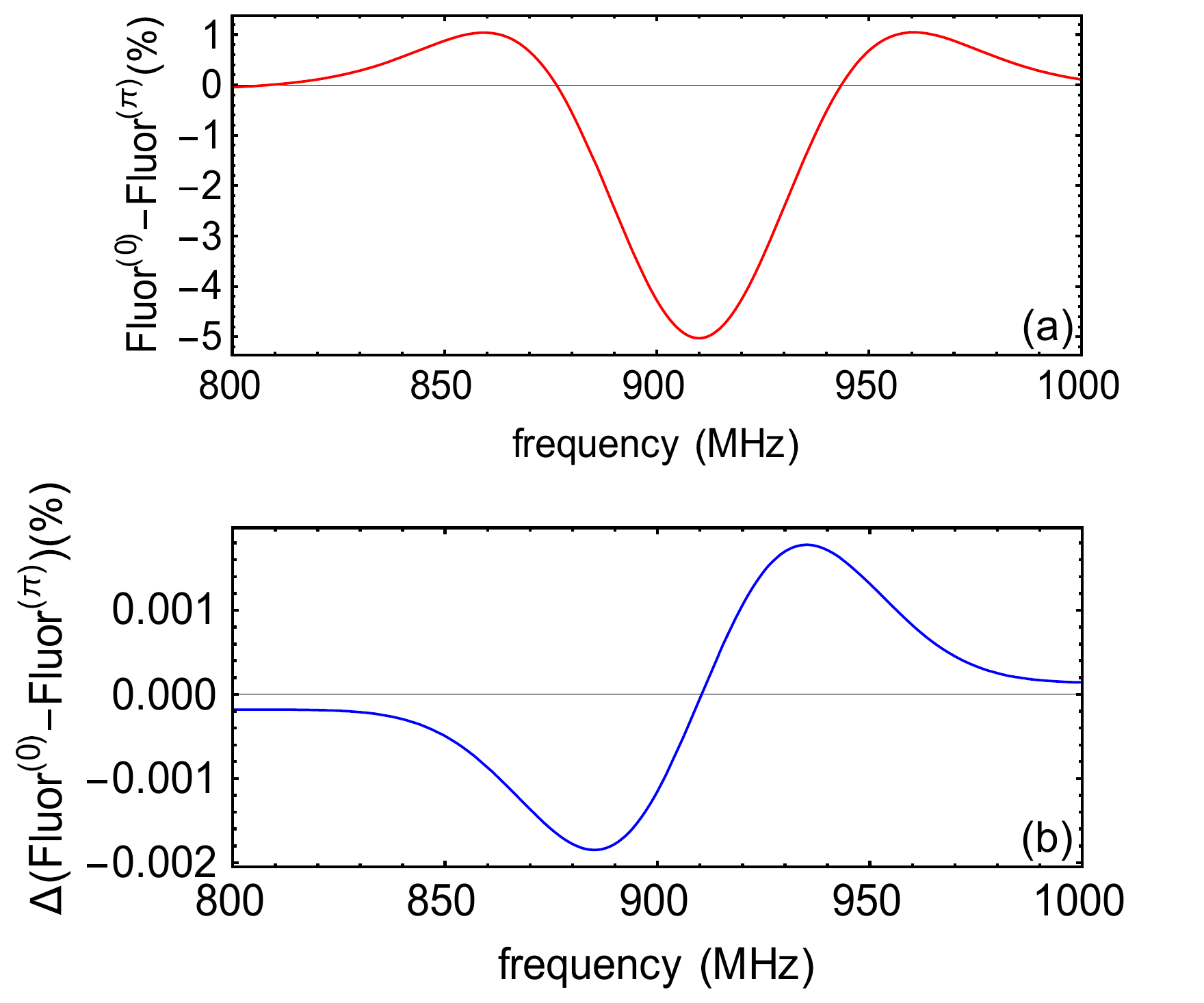}
\caption{\label{fig:Hshapes} (Color online) 
Line shapes for an SOF 
2S$_{1/2}$-to-2P$_{1/2}$
atomic hydrogen
Lamb-shift measurement.
The full line shape for 
the 
Lyman-$\alpha$ 
fluorescence predicted for an 
initial population of 
1.0
divided equally among the four
2S$_{1/2}$ 
states
and a population of 
0.34
divided equally among the four 
3S$_{1/2}$ 
states
is shown in 
(a).
This line shape includes a small
distortion due to the 
fact that coherences
between 
3S 
and 
3P
states 
can radiatively decay into 
coherences
between 
2S 
and 
2P
states.
The difference between
the calculations with and without this
decay of coherences for the 
SOF 
measurement
are shown in 
(b).
The dispersion in (b) is due to the same effect as
that shown in 
Fig.~\ref{fig:SOFLineShapes}(c).
The distortion in (b) causes a shift of
approximately 
-9.5~kHz.
}
\end{figure} 

The calculation is computationally intensive and 
employed the SHARCNET computer cluster.
This calculation is far beyond that which could 
have been performed at the time 
of the measurement of 
Ref.~\cite{PRL46.232}.
The SOF lineshape 
obtained from integrating the equations is 
shown in 
Fig.~\ref{fig:Hshapes}(a).
Part 
(b) 
of the figure shows the 
difference between the  
calculation of the 
SOF
line shape
with and without
the terms analogous to 
the 
square-root 
term in 
Eq.~(\ref{eq:4LevelEqs}c),
that is,
with and without allowing 
coherences in the 
$n$=3
states to radiatively decay to 
coherences in the 
$n$=2
states. 
The difference reveals a dispersion
line shape,
similar to that of 
Fig.~\ref{fig:SOFLineShapes}(c).
This dispersion curve results from the
sum of all of the interferences 
between 
2S-to-2P
and 
3S-to-3P 
resonances,
with the sign of each interference 
determined by the relative signs of
the matrix elements of the 
3S-to-2P
and 
3P-to-2S
decay paths.
The shift caused by the presence of 
3S
atoms is dominated by the
dispersion line shape due to 
radiative decay of coherences
between
3S
states and 
3P 
states 
(which decay into 
coherences between 
2S
and 
2P 
states).
Using the set of  
frequencies of 
Ref. \cite{LundeenMetrologia},
and using the 
symmetric-points 
method described in that work,
the calculated shift due to the 
3S
atoms is
-9.5~kHz.
A 
+9.5~kHz 
correction would have to be
applied to the measurement of
Ref.~\cite{PRL46.232} 
to compensate for this shift if 
only the 
3S
state were present,
and if its
population relative to the 
2S
state
were 0.34.
However, 
to calculate the full shift, 
a similar analysis would also have
to be performed for 
higher-$n$ 
states,
and precise knowledge of the populations in 
each of the 
$n>2$ 
states would be necessary.
It is not possible to infer these populations 
to a sufficient level of accuracy from 
Refs.~\cite{PRL46.232}
and \cite{LundeenMetrologia}.
The calculation presented here does,
however, 
conclude that the correction 
can be expected to be on the 
10-kHz 
scale.

The calculation presented in this section
is far more complicated 
than the simple 
four-level
model of 
Section~\ref{SOF},
but the physics dominating the calculated
shifts 
remains the same.
The parameters used for   
Figs.~\ref{fig:RabiLineShapes}
and
\ref{fig:SOFLineShapes} 
correspond to the 
four-level system consisting of 
$|2{\rm S}_{1/2},f\!\!=\!\!0,m_f\!\!=\!\!0\rangle$,
$|2{\rm P}_{1/2},f\!\!=\!\!1,m_f\!\!=\!\!0\rangle$,
$|3{\rm S}_{1/2},f\!\!=\!\!0,m_f\!\!=\!\!0\rangle$,
and
$|3{\rm P}_{3/2},f\!\!=\!\!1,m_f\!\!=\!\!0\rangle$.
The full calculation extends this 
four-level
model to include all of the 
states for 
$n$=1, 
2 and 3,
which allows for more decay channels, 
for more coherences,
and for more interfering resonances.

The 
10-kHz 
scale of this interference shift 
makes it important, 
since the SOF measurement of 
Ref.~\cite{PRL46.232}
has a precision of 
9~kHz,
and the discrepancy between
this measurement and the interval calculated
\cite{horbatsch2016tabulation}
using the new,
more accurate measurement
of the proton charge radius from
muonic hydrogen
\cite{
pohl2010size,
antognini2013proton}
is
13~kHz.
The ongoing
Lamb-shift 
measurement by our 
group has a precision goal of a few kHz,
and,
therefore, 
these shifts will have to be considered very carefully.

\section{summary}

We have identified a new systematic effect important
to precision measurements. 
The effect allows for an interference between two 
resonant transitions that do not have a common 
initial nor final state, 
but,
rather,
have states connected by radiative decay.
The interference causes a 
line-shape
distortion and 
a shift for the precision measurement.
We find that this effect is large enough 
to have an important impact on 
the existing
\cite{PRL46.232}
and ongoing 
measurements of the 
hydrogen
$n$=2
Lamb shift.
As such, 
they may contribute to the 
current discrepancy 
\cite{
bernauer2014proton,
pohl2013muonic,
carlson2015proton}
between
muonic and electronic measurements
of the proton radius.

\section{acknowledgements}
This work is supported by NSERC and CRC
and used SHARCNET for computation.

\bibliography{LunPipInterference}

\end{document}